\documentclass[12pt]{article}

\usepackage{comment}
\usepackage{tikz}

\usepackage{mathtools}

\usepackage[mathscr]{euscript}

\usepackage{slashed}

\usepackage{amsmath}

\date{}

\title{\textbf{
First-Order Gauge Invariant Generalization of the Quantum Rigid Rotor
}}
\author{ \textbf{
Suzicleide L. de Oliveira, Camila M. B. Santos, Ronaldo Thibes}
\\\\
\textit{\small{Departamento de Ci\^encias Exatas e Naturais}},\\
\textit{\small{Universidade Estadual do Sudoeste da Bahia}},\\
\textit{\small{Rodovia BR 415, km 03, s/n - Itapetinga - BA}},\\
\textit{\small{45700-000 Brazil}}
 }

\usepackage{graphicx}
\usepackage{amssymb,amsfonts,amssymb,amsthm}
\usepackage{amsmath}

\begin{document}

\maketitle

\abstract{A first-order gauge invariant formulation for the two-dimensional quantum rigid rotor is long known in the theoretical physics community as an isolated peculiar model.
Parallel to that fact,
the longstanding constraints abelianization problem, aiming at the conversion from second to first class systems for quantization purposes, has been approached a number of times in the literature with a handful of different forms and techniques and still continues to be a source of lively and interesting discussions.
Connecting these two points,
we develop a new systematic method for converting second class systems to first class ones,
valid for a class of systems encompassing the quantum rigid rotor as a special case. In particular the gauge invariance of the quantum rigid rotor is fully clarified and generalized in the context of arbitrary translations along the radial momentum direction. Our method differs substantially from previous ones as it does not rely neither on the introduction of new auxiliary variables nor on the {\it a priori} interpretation of the second class constraints as coming from a gauge-fixing process.
}

\section{Introduction}
When it comes to fundamental physics,
gauge invariance has been one of the main cornerstones for the most important models in quantum field theory.  Since the enormous success of quantum electrodynamics (QED), quantum chromodynamics (QCD) and the standard model as a whole, gauge symmetry has acquired the status of an essential guiding principle  in the quest for paramount theories describing nature.  Attesting this fact, it is notorious that all known serious proposals for a quantum gravity theory, including (super)strings, enjoy gauge symmetry.  In this way, one important quantization path for second class dynamical systems stems from the idea of converting the second class constraints to first class ones.  Indeed the latter are responsible for generating gauge transformations and are always present in the context of gauge symmetry.

On the other hand, a necessary downside for gauge invariance lies in the fact that we must use more field variables than the actual number of degrees of freedom.  In a natural way, the dynamics of any gauge system 
contains a set of constraint conditions among the field variables, being thus described by a Hamiltonian {\it constrained} theory.  In general the elimination of some of the variables in favor of others by means of enforcing the constraint equations is an artificial one, breaking important physical symmetries, and usually presents heavy technical difficulties.  Instead, the quantization of such systems can be done, either by canonical or path integral methods, by maintaining and treating properly the constraints at quantum level.

A complete algorithm for unraveling all constraints of a given dynamical system in a systematic way, as well as the classification scheme into first and second class ones, has been introduced long ago by Dirac \cite{Dirac:1950pj} and Anderson and Bergmann \cite{Anderson:1951ta} and is described in the classical texts
\cite{Dirac, Hanson:1976cn, Sundermeyer:1982gv}.
Modern reviews as well as later developments of this important constraint classification program, nowadays known as the Dirac-Bergmann algorithm, can be seen for instance in
\cite{Gitman:1990qh, Henneaux:1992ig, Rothe:2010dzf}.
Basically we say that a constraint is second class when it has a non-null Poisson bracket in the phase space constraint hypersurface with at least one of the other remaining constraints.
One important key feature of Dirac's constraint classification relies on the fact that
only first class constraints are related to gauge symmetries.
This naturally raises the question of whether it could be possible to contextualize or modify a given model in such a way that the second class constraints would behave as first class.

This attractive idea of converting second class constraints to first class in order to produce gauge symmetries has led to a plethora of methods in the 
theoretical physics literature.  We may mention for instance the introduction
of auxiliary variables in phase space \cite{Stueckelberg:1957zz, Wess:1971yu,Batalin:1986aq, Batalin:1986fm, Batalin:1989dm, Fujiwara:1989ia, Batalin:1991jm, Banerjee:1993pm, Amorim:1994ft, Natividade:2000sh, Abreu:2016bie},
the interpretation of the second class constraints as gauge-fixing conditions coming from a larger theory \cite{Harada:1988aj, Mitra:1990mp, Mitra:1990qt,Anishetty:1992yk, Vytheeswaran:1994np, Neto:2006gt, Neto:2009rm}
or the use of first-order Lagrangians  \cite{Amorim:1995sh, Amorim:1999xr}.
In the first case, by introducing extra auxiliary variables and extending the phase space, it is possible to have room for cancellations in the constraint algebra in order that originally second class constraints become first class.  A successful systematic application of this method came to be known as the Batalin-Fradkin-Fradkina-Tyutin (BFFT) formalism \cite{Batalin:1986aq, Batalin:1986fm, Batalin:1989dm}.
Alternatively, the second approach above mentioned tries to reverse the gauge-fixing process assuming that a second class system can be thought as coming from an original gauge invariant one where the gauge freedom has been fixed by imposing subsidiary conditions which appear in the same level as the constraints.  This interpretation, first considered by Harada, Mukaida, Mitra and Rajaraman \cite{Harada:1988aj,Mitra:1990mp,Mitra:1990qt}, has led to the so-called gauge unfixing method \cite{Anishetty:1992yk, Vytheeswaran:1994np} which has been recently generalized by Jorge Ananias Neto \cite{Neto:2006gt, Neto:2009rm}.

The methods above rely all in the Hamiltonian formalism.  In our current work however, understanding that gauge symmetry is more naturally realized starting from a Lagrangian function, we take advantage of a first-order Lagrangian \cite{Amorim:1995sh, Amorim:1999xr, Hojman:1981sf, Carinena:1988zj,Faddeev:1988qp, BarcelosNeto:1991kw} to forthright produce gauge symmetry.
Inspired in a insightful paper of Nemeschansky, Preitschopf and Weinstein \cite{Nemeschansky:1987xb} and its recent generalization in \cite{Barbosa:2018bng, Barbosa:2018dmb}, we introduce a suitable first-order Lagrangian based on an invariant potential which changes the second class constraint structure of a given second class system to first class generating gauge symmetry.  As a matter of fact, in reference
\cite{Nemeschansky:1987xb}, the authors considered a simple model describing a particle in a circular path with gauge symmetry and discussed its quantization and BRST symmetry at quantum level.  Their original goal actually was to use a mechanical model in order to trace analogies with quantum field theory with respect to gauge and BRST symmetries and ghost fields.  That model was subsequently used in several papers from which we cite \cite{Gupta:2009dy, Shukla:2014hea, Shukla:2014spa}.  In reference \cite{Gupta:2009dy}, Gupta and Malik applied the superfield approach to the gauge invariant two-dimensional rigid rotor as presented in \cite{Nemeschansky:1987xb}, constructing a toy model for the Hodge theory.  The quantum BRST symmetries of this model were then deeper scrutinized in \cite{Shukla:2014hea, Shukla:2014spa}.
But somehow the inherent circular symmetry seemed to be always mandatory.  Finally last year, the couple of references \cite{Barbosa:2018bng, Barbosa:2018dmb} generalized further that circular model considering a particle moving in a generic conic path and showing the accidental role of the mentioned circular symmetry.
In the present letter we push it further and show that the target system may be defined in arbitrary dimensions and the original second class constraint may be given by a broader differential function of the generalized coordinates encompassing all previous known cases.
We also introduce a non-degenerate two-form in the kinetic term. 
Starting from this more general second class initial system we show how to obtain the corresponding first class version with gauge symmetry.

This letter is organized as follows.
In Section {\bf 2} below we define our working constrained system and using the Dirac-Bergmann algorithm obtain the whole constraint structure and compute its Dirac brackets.
The initial system is shown to possess only second class constraints and therefore no gauge freedom.  In Section {\bf 3} we construct an equivalent gauge invariant system, exhibiting the explicit local symmetry.  We explore the singularity of the constraint matrix to obtain the first class constraints.  In Section {\bf 4} we consider two specific applications, namely, the rigid rotor and a particle on an elliptical path.  In the first case we show how our formulation reproduces \cite{Nemeschansky:1987xb}. The elliptical path is implemented by using a square root in the constraint equation which also constitutes a novelty.  We close in Section {\bf 5} with our conclusion and final remarks.

\section{The Starting Second Class System}
Given a symmetric invertible square matrix $f_{ij}$ as function of the variables $q_k$, for $i, j, k = 1, \dots, n$, and a thrice differentiable function $T(q_k)$, we define a prototypical second class system given by the Lagrangian\footnote{For simplicity we are considering mechanical systems.  The description of fields can in principle be achieved by allowing the discrete indexes to take continuous values through a limiting process.}
\begin{equation}\label{L}
L(q_0,q_k,{\dot{q}}_k)=\frac{m}{2}\sum_{i,j=1}^{n}f_{ij}^{-1}(q_k){{\dot{q}}_i}{{\dot{q}}_j}
+q_0T(q_k)
\end{equation}
depending on the $2n+1$ variables $q_0,q_k,{\dot{q}}_k$.
These latter in turn, as usual, are considered to depend on a real parameter $t$ representing physical time and the dot denotes the derivative with respect to it.
In this way, $(q_0,q_k,\dot{q_0},\dot{q}_k)$ characterize the configuration space where the system evolves dynamically.
The global mass parameter $m$ will be useful to keep track of physical dimensions in the coming expressions.
The variable $q_0$ does not show up in the kinetic term and actually plays the role of a Lagrange multiplier enforcing the constraint $T(q_k)=0$.  This grants the system a total of $n-1$ physical degrees of freedom.

A systematic Dirac-Bergmann constraint analysis of system (\ref{L}) shows that we have in fact four second class constraints.
This can be seen as follows.
First of all,
by introducing conjugate momenta variables $p_0, p_k$, the corresponding canonical Hamiltonian in phase space reads
\begin{equation}\label{H}
H=\frac{1}{2m}\sum_{i,j=1}^{n}f_{ij}(q_k)p_ip_j-q_0 T(q_k)\,.
\end{equation}
Then, since the Lagrangian (\ref{L}) does not depend on $\dot{q}_0$, we have a first trivial primary constraint $\chi_1 \equiv p_0$ whose time conservation under the Hamiltonian produces the desired relation $\chi_2 \equiv T(q_k)$ as a secondary constraint.
Next, proceeding with the Dirac-Bergmann algorithm,
a straightforward calculation shows that further conservation of $\chi_2$ leads to a couple more constraint relations, namely\footnote{We are assuming that the given function $T(q_k)$ satisfies
\begin{equation}\nonumber
\frac{\partial T(q_k)}{\partial q_i}\neq 0
\end{equation}
for each $i=1,\dots,n$.},
\begin{equation}\label{chi3-p}
\chi_3 \equiv \frac{1}{m}\sum_{i,j=1}^{n}f_{ij}p_i \frac{\partial T}{\partial q_j}
\end{equation}
and
\begin{equation}\label{chi4-p}
\chi_4 \equiv \frac{1}{m^2}\sum_{i,j=1}^{n} Q_{ij}p_i p_j 
+\frac{q_0}{m}\sum_{i,j=1}^{n} f_{ij}\frac{\partial T}{\partial q_i}
\frac{\partial T}{\partial q_j}
\,,
\end{equation}
with
\begin{equation}
Q_{ij}\equiv
\sum_{k, l=1}^{n}
\left\{
\left(
\frac{\partial f_{ik}}{\partial q_l}\frac{\partial T}{\partial q_k}
+
{f_{ik}}\frac{\partial^2 T}{\partial q_k\partial q_l}
\right)f_{lj}
-\frac{1}{2}
f_{kl}\frac{\partial f_{ij}}{\partial q_k}\frac{\partial T}{\partial q_l}
\right\}
\,.
\end{equation}
For computational convenience and a more easily handling of formulae we introduce at this point the condensed brief notation
\begin{equation}\label{bnot}
T_i\equiv\frac{\partial T}{\partial q_i}\,,~~~~~T_{ij}\equiv\frac{\partial^2 T}{\partial q_j \partial q_i}~~~~~\mbox{etc}
\end{equation}
and
\begin{equation}
f_{ij,k}\equiv \frac{\partial f_{ij}}{\partial q_k}
\,,~~~~~
f_{ij,kl}\equiv \frac{\partial^2 f_{ij}}{\partial q_k\partial q_l}
\,,~~~~~
\mbox{etc}
\end{equation}
for the partial derivatives\footnote{We assume the order of the partial derivatives always commute, i.e., $T_{ij}=T_{ji}$ etc.} with respect to the coordinates $q_i$. Furthermore, from now on, we also use Einstein's repeated index convention sum  from $1$ to $n$ for Latin indexes $i, j, k, l$.  In this way the two previous constraints can be compactly rewritten as
\begin{equation}\label{chi3}
\chi_3 = \frac{f_{ij} T_ip_j}{m}
\end{equation}
and
\begin{equation}\label{chi4}
\chi_4 = \frac{1}{m^2}Q_{ij}p_ip_j+\frac{q_0}{m}f_{ij}T_iT_j
\,,
\end{equation}
with $Q_{ij}$ being simpler expressed as
\begin{equation}\label{Qij}
Q_{ij}\equiv
\left(
f_{ik,l}T_k+f_{ik}T_{kl}
\right)f_{lj}
-\frac{1}{2}f_{ij,k}f_{kl}T_l
\,.
\end{equation}
Note that, precisely because of Einstein's convention, we do not need to bother explicitly writing the summation symbol anymore.
Finally,
the fact that the complete set of constraints $\chi_r$, with $r=1,\dots,4$, is second class can be seen by investigating the constraint matrix
$C_{rs}=[\chi_r,\chi_s]$ given explicitly by
\begin{equation}\label{CM}
C_{rs}
=
\displaystyle
\left[
\begin{array}{cccc}
0 & 0 & 0 &-f_{ij}T_i T_j/m \\
0 & 0 &f_{ij}T_i T_j/m &2Q_{ij}T_ip_j/m^2 \\
0 & -f_{ij}T_i T_j/m  & 0 & R \\
f_{ij}T_i T_j/m  & -2Q_{ij}T_ip_j/m^2 & -R & 0
\end{array}
\right]
\end{equation}
with
\begin{equation}
\displaystyle
R\equiv \frac{1}{m^3}
\left[
2Q_{ik}\left( f_{jl,i}T_j+f_{jl}T_{ij} \right)
-Q_{kl,i}f_{ij}T_j  
\right]p_kp_l
-\frac{q_0}{m^2}
\left(
2f_{ik}f_{jl}T_{ij}+f_{ij}T_if_{kl,j}
\right)T_kT_l
\,.
\end{equation}
Since we are assuming $f_{ij}$ non-degenerated and $T_i\neq 0$, the matrix $C_{rs}$
is clearly nonsingular with determinant given by
\begin{equation}
\det C_{rs} = \left[ f_{ij}T_i T_j/m \right]^4
\,.
\end{equation}
Therefore we have just shown that the Lagrangian (\ref{L}) describes a genuine second class system and in principle does not enjoy gauge invariance at all.

As is well-known, for a constrained system, the Poisson brackets algebra for the phase
space variables does not concur properly with the constraints and hence is not suitable for a consistent quantization.
For that purpose we need to compute instead the Dirac brackets which are defined for two
arbitrary phase space functions $F$ and $G$ as
\begin{equation}\label{DB}
[F,G]=[F,G]_{PB}-\sum_{r,s=1}^4[F,\chi_r]_{PB}\,C^{rs}\,[\chi_s,G]_{PB}
\end{equation}
where $C^{rs}$ denotes the inverse of the constraint matrix (\ref{CM}) and the subscript $PB$ stands for the usual Poisson bracket.
In this way, by inverting $C_{rs}$ and using (\ref{DB}), the Dirac brackets among all phase variables can be straightforwardly computed leading to the non-null results
\begin{equation}\label{qipj}
[q_i, p_j] = \frac{(\delta_{ij}f_{kl}-f_{ik}\delta_{lj})T_kT_l}{f_{kl}T_kT_l}
\,,
\end{equation}
\begin{equation}\label{pipj}
[p_i, p_j] = \frac{(f_{kl,i}T_j-f_{kl,j}T_i)T_k
+f_{kl}(T_{ki}T_j-T_{kj}T_i)}{f_{kl}T_kT_l}\,p_l
\,,
\end{equation}
\begin{equation}\label{q0qi}
[q_0, q_i] = \frac{2(Q_{ij}f_{kl}-f_{ik}Q_{jl})T_kT_lp_j}{mf_{kl}T_kT_l}
\,,
\end{equation}
and
\begin{eqnarray}\label{q0pi}
[q_0, p_\kappa] &=& \frac{p_kp_l}{m(f_{kl}T_kT_l)^2}
\displaystyle
\left\{
2Q_{ik}
\left[
T_i(f_{jl,\kappa}T_j+f_{jl}T_{j\kappa})
-T_\kappa(f_{jl,i}T_j+f_{jl}T_{ij})
\right]
\nonumber\right.\\&&\left.
+f_{ij}T_j(Q_{kl,i}T_\kappa-Q_{kl,\kappa}T_i)
\right\}
+\frac{q_0f_{ij}T_jT_l}{(f_{kl}T_kT_l)^2}
\left[
2f_{kl}(T_{ik}T_\kappa-T_{\kappa k}T_i)  
\nonumber\right.\\&&\left.
+T_k(f_{kl,i}T_\kappa - f_{kl,\kappa}T_i)
\right]
\,,
\end{eqnarray}
for $i, j, k, l, \kappa =1, \dots , n$.
The canonical quantization of this system can be performed by promoting the phase space variables to operators satisfying commutation relations given by the Dirac brackets above and acting on a Hilbert space of complex functions of $(q_0,q_i,p_0,p_i)$.  Issues related to operator ordering ambiguities can be tackled for instance by requiring Hermicity but may depend on the specific details of the model.  In the present letter however, we are chiefly concerned with obtaining gauge symmetry as outlined in the Introduction.  In order to produce gauge invariance, in the next section we shall describe the same system (\ref{L}) by an equivalent
first-order Lagrangian exhibiting first class constraints.

\section{The Gauge Invariant System}
In this section we introduce a first-order gauge invariant system equivalent to the second class Lagrangian (\ref{L}).  Following and generalizing the ideas of \cite{Barbosa:2018bng, Barbosa:2018dmb} we consider the potential
\begin{equation}\label{W}
W(q_k,p_k)=\frac{F_{ijkl}T_iT_jp_kp_l}{{2m}f_{ij}T_iT_j}
\end{equation}
with
\begin{equation}\label{Fijkl}
F_{ijkl}\equiv f_{ij}f_{kl} - f_{ik}f_{jl}
\,.
\end{equation}
The tensor $F_{ijkl}$ defined in (\ref{Fijkl}) behaves under index
exchanges as
\begin{equation}
F_{ijkl} = F_{jilk} = F_{klij}
\end{equation}
and
\begin{equation}\label{anti}
F_{ijkl} = -F_{ikjl} = -F_{ljki}
\,.
\end{equation}
Concerning the antisymmetric property (\ref{anti}) of $F_{ijkl}$, we see that if $\Omega(t)$ denotes an arbitrary time-dependent function, the transformation $p_i\rightarrow p_i+\Omega T_i$ leaves the potential $W(q_k,p_k)$ invariant.  This permits us to construct the first-order Lagrangian
\begin{equation}\label{Lfo}
L_{fo}=p_i\dot{q}_i 
-
W(q_k,p_k)
+q_0T(q_k)
\end{equation}
which 
varies under the local transformation
\begin{equation}\label{gs}
\begin{cases}
~~p_i\longrightarrow p_i+\Omega T_i\,, \\
~~q_0\longrightarrow q_0 + \dot{\Omega} \,,
\end{cases}
\end{equation}
as a total time derivative
\begin{equation}
L_{fo}\longrightarrow L_{fo} +\frac{d}{dt}(\Omega T)
\end{equation}
leading thus to a gauge symmetry of the corresponding action.
By applying the Dirac-Bergmann algorithm, we show in the remaining of this section that the constraint content of systems
(\ref{L}) and (\ref{Lfo}) is equivalent.  In particular, although only the latter enjoys gauge invariance, we shall soon see that the net number of degrees of freedom of both systems is exactly the same.

Considering
(\ref{Lfo}) as an ordinary Lagrangian system depending on the
$2n+1$ variables $(q_0,q_i,p_i)$ we introduce the corresponding
canonical
momenta $(p_0,P_i,\Pi_i)$ and obtain the $2n+1$ primary constraints
\begin{equation}\label{primaryfo}
\begin{cases}
\Phi_0 = p_0\,,\\
\Phi_{1i} = P_i - p_i \,, \\
\Phi_{2i} = \Pi_i \,.
\end{cases}
\end{equation}
Then a Legendre transformation to phase space produces immediately the expected Hamiltonian
\begin{equation}
H=
W(q_k,p_k)
-q_0T(q_k)
\end{equation}
and time conservation of the primary constraints (\ref{primaryfo}) leads to the secondary one
\begin{equation}\label{secondaryfo}
\Phi=T(q_k)
\end{equation}
which happens to be the very main constraint of the initial system.
By computing the Poisson brackets among the relations (\ref{primaryfo}) and (\ref{secondaryfo}) we construct the constraint matrix
\begin{equation}\label{CMfo}
C_{rs}
=
\left[
\begin{array}{cccc}
0 & 0_j & 0_j &0 \\
0_i & 0_{ij} &-\delta_{ij} &-T_i \\
0_i &\delta_{ij}  & 0_{ij} & 0_i \\
0 & T_j & 0_j & 0
\end{array}
\right]
\end{equation}
with $r, s = 1,\dots,2n+2$ and $i,j=1,\dots,n$.  Just to be clear, in terms of notation, we remark that equation (\ref{CMfo}) above denotes a $(2n+2)\times (2n+2)$ square antisymmetric matrix within which $0_j$ represents the $1\times n$ null row matrix, $0_i$ the $n\times1$ column matrix and $0_{ij}$ the $n\times n$ square null matrix. We can see that now, contrary to the previous case, the constraint matrix  $C_{rs}$ is singular.
Not only are the first row and column of (\ref{CMfo}) null but actually its rank is $2n$.
In fact, in addition to
\begin{equation}
v_0\equiv
\displaystyle\left[
\begin{array}{cccc}
1 & 0_i & 0_i & 0
\end{array}
\right]
\,
\end{equation}
it follows that
\begin{equation}
v\equiv
\displaystyle\left[
\begin{array}{cccc}
0 & 0_i & T_i & -1
\end{array}
\right]
\end{equation}
is also a null mode of $C_{rs}$.  This signals the presence of two
first class constraints in the theory.  Besides $\Phi_0 $, which has
a null Poisson bracket with all other ones,
we can check that the constraints linear combination
\begin{equation}
\tilde{\Phi}\equiv \Phi - T_i\Phi_{2i}
\end{equation}
is also first class.  Indeed we have 
\begin{equation}
[\tilde{\Phi},\Phi_{1i}]_{PB} = -T_{ij}\Phi_{2j}
\end{equation}
and
\begin{equation}
[\tilde{\Phi},\Phi_{0}]_{PB} = [\tilde{\Phi},\Phi_{2i}]_{PB} = 0
\,.
\end{equation}
Therefore the constraint content of system (\ref{Lfo}) consists of
$2n$ second class and two first class constraints.  The
second class ones are trivial in a certain sense, stating the clearly obvious fact that since the starting Lagrangian is first-order, the configuration space variables $(q_i,p_i)$ behave as canonically conjugate pairs.  This is very similar to what happens when one applies the Dirac-Bergmann algorithm to the first-order Dirac Lagrangian in field theory in which trivial second class constraints relate $\psi$ and $\bar\psi$ as a pair of canonically conjugate fields.  On the other hand, the two additional first class constraints are responsible for the gauge symmetry (\ref{gs}).  The number of degrees of freedom (DOF) of a constrained dynamical system is given by \cite{Henneaux:1992ig}
\begin{equation}
\mbox{DOF}= \frac{2N-S-2F}{2}
\end{equation}
where $F$ and $S$ denote respectively the number of first and second class constraints and $N$ the total number of variables in the configuration space.  In our present case, for the first-order Lagrangian (\ref{Lfo}),  we have $N=2n+1$, $F=2$ and $S=2n$ leading to a net result of $n-1$ degrees of freedom, exactly the same as the original second class model (\ref{L}).
We have thus achieved our goal of producing a first class system equivalent to the starting Lagrangian (\ref{L}) enjoying gauge symmetry.

In the next section we work two particular elucidating examples of the prototypical Lagrangian (\ref{L}) computing its Dirac brackets and obtaining the corresponding gauge invariant system of the form (\ref{Lfo}).
\section{Examples}
In order to illustrate the general ideas discussed in the previous sections, in the following we work out two specific examples, namely, the quantum rigid rotor and a particle moving in an elliptical path implemented with a square-root-type constraint.
In both cases we start with a second class Lagrangian and obtain a corresponding gauge invariant one.
\subsection*{The 2D Quantum Rigid Rotor}
The two-dimensional classical rigid rotor consists of a mass $m$ particle constrained to move along a radius $r$ circle which can be realized by
the simple Lagrangian
\begin{equation}\label{Lrr}
{L} = \frac{m}{2}\left( \dot{r}^2 +r^2\dot{\theta}^2 \right) + z(r-a)
\end{equation}
where $r$ and $\theta$ denote the radial and angular particle position variables and $z$ a Lagrange multiplier.
The quantum version can be obtained by promoting the variables $r$, $\theta$ and $z$ to operators satisfying commutation relations which generalize Heisenberg's uncertainty principle.  However, due to the singular nature of (\ref{Lrr}), instead of the Poisson brackets, the Dirac ones should be sent to the quantum commutators.
A complete quantum implementation along these lines can be seen for instance in \cite{Scardicchio:2002}. 
Alternatively,
back in 1988, Nemeschansky, Preitschopf and Weinstein \cite{Nemeschansky:1987xb}
introduced a gauge invariant version of the two-dimensional rigid rotor (\ref{Lrr}) by constructing an {\it ad hoc} reduced Hamiltonian for a constrained system.  In their treatment of the rigid rotor quantization problem in \cite{Nemeschansky:1987xb}, the authors give a prescription of ``simply throwing away'' a piece of the Hamiltonian, namely, the one which does not commute with the constraints\footnote{That prescription was actually originally discussed by Dirac in \cite{Dirac}.}. Here we show how to reobtain their result in a systematic and clear way as a special case of our general framework.

Let us first confirm that the Lagrangian (\ref{Lrr}) corresponds to a consistent second class system of the form (\ref{L}).  First of all, by direct comparison we see that we have $n=2$ and $f_{ij}(q_k)$ and $T(q_k)$ given respectively by
\begin{equation}\label{frtheta}
f(r,\theta)=
\left[ \begin{array}{cc}
 1 &0    \\ 
 0 & 1/r^2 \\
 \end{array} \right]
\,,
\end{equation}
and
\begin{equation}
T(r,\theta) = r - a
\,,
\end{equation}
with $q_0\equiv z$, $q_1\equiv r$ and $q_2\equiv\theta$.
In particular, concerning the partial derivatives of $T(r,\theta)$ above
in the notation of equation (\ref{bnot})
we have
\begin{equation}\label{TrTtheta}
T_1\equiv T_r = 1\mbox{~and~}T_2\equiv T_\theta = 0
\,.
\end{equation}
Next,
passing to phase space, we introduce the three canonically conjugate
momenta $p_r$, $p_\theta$, $p_z$ from which immediately follows the primary constraint
\begin{equation}
\chi_1 \equiv p_z
\,.
\end{equation}
Then,
following the Dirac-Bergmann algorithm, we compute the
canonical Hamiltonian
\begin{equation}
H=\frac{p_\theta^2}{2mr^2}+\frac{p_r^2}{2m} -z(r-a)
\,,
\end{equation}
and impose time conservation of the primary constraint
$\chi_1$.  This leads to the desired
circular path constraint
\begin{equation}
\chi_2=T(r,\theta) = r - a
\end{equation}
and, corresponding to equations (\ref{chi3}) and (\ref{chi4})
to the second class constraints
\begin{equation}
\chi_3 = f_{rr}T_rp_r + f_{\theta\theta}T_\theta p_\theta = \frac{p_r}{m}
\end{equation}
and
\begin{equation}
\chi_4 = \frac{1}{m^2}Q_{\theta\theta}p_\theta^2+\frac{z}{m}(T_r^2+\frac{1}{m^2}T_\theta^2) = \frac{p_\theta^2}{m^2 r^3} + \frac{z}{m} 
\,.
\end{equation}
To obtain $\chi_4$ above, besides (\ref{TrTtheta}), we have used the fact that, from equation (\ref{Qij}), we have
\begin{equation}
Q_{ij}=
\left[ \begin{array}{cc}
 0 &0    \\ 
 0 & 1/r^3 \\
 \end{array} \right]
\,.
\end{equation}
Furthermore, in the present case,
\begin{equation}
Q_{ij}T_i p_j = Q_{rr}T_rp_r + Q_{\theta\theta}T_\theta p_\theta = 0
\end{equation}
and the constraint matrix (\ref{CM}) can be checked to be given by
\begin{equation}
C_{rs}=
\left[ \begin{array}{cccc}
 0 &0  &0  &-1/m   \\ 
 0 & 0 &  1/m & 0\\
 0 & -1/m & 0  & \frac{3p_\theta^2}{m^3r^4}  \\
 1/m & 0 &  -\frac{3p_\theta^2}{m^3r^4}& 0
 \end{array} \right]
\,.
\end{equation}
As anticipated, the matrix $C_{rs}$ is nonsingular since it satisfies
\begin{equation}
\det C_{rs}=
1/m^4
\,,
\end{equation}
which ensures the second class nature of the complete constraints set $\chi_r$, $r=1, \dots,4$.
The non-null Dirac brackets among the phase space variables, according to the general relations
(\ref{qipj}) to (\ref{q0pi}), are given by
\begin{equation}
[
\theta, p_\theta
] = 1
\mbox{~~~and~~~}
[z,\theta] = - \frac{2p_\theta}{mr^3}
\,.
\end{equation}
The thorough canonical quantization of this system using this Dirac brackets algebra, without gauge freedom, can be found in \cite{Scardicchio:2002}.

After confirming the second class nature of (\ref{Lrr}) and computing the Dirac brackets, our next step is to construct the corresponding gauge invariant model.  By plugging (\ref{frtheta})
and (\ref{TrTtheta}) into (\ref{W}) and (\ref{Fijkl}) we obtain the potential
\begin{equation}
W(r,p_\theta)= \frac{p_\theta^2}{2mr^2}
\end{equation}
and form the first-order gauge invariant Lagrangian defined in (\ref{Lfo}) as
\begin{equation}\label{Linvrtheta}
L_{fo} =  p_r \dot{r} + p_\theta \dot{\theta} - \frac{p_\theta^2}{2mr^2} + z(r-a)
\,.
\end{equation}
This is precisely the Lagrangian obtained in reference \cite{Nemeschansky:1987xb} and used in the later works \cite{Gupta:2009dy, Shukla:2014hea, Shukla:2014spa}.
We can see by inspection that (\ref{Linvrtheta}) transforms as a total derivative under
\begin{equation}\label{gtransf}
\begin{cases}
p_r\rightarrow p_r+\Omega\,,\\
z\rightarrow \dot{\Omega}\,,
\end{cases}
\end{equation}
for an arbitrary local function $\Omega$.
In other words, applying (\ref{gtransf})to $L_{fo}$ in (\ref{Linvrtheta}) we have
\begin{equation}
L_{fo}\rightarrow L_{fo} +\frac{d}{dt}\left[ \Omega (r-a) \right]
\end{equation}
leading to a gauge symmetry of the action.
The constraint structure associated to the system defined by (\ref{Linvrtheta}) can be analyzed by introducing the momenta variables $(p_z, P_r, P_\theta, \Pi_r, \Pi_\theta)$, respectively canonically conjugated to $(z, r, \theta, p_r, p_\theta)$. The momenta definition, in terms of the derivatives of the Lagrangian with respect to the velocities, immediately produces five primary constraints corresponding to (\ref{Linvrtheta}) which we denote here by
\begin{equation}\label{primaryconsts}
\begin{cases}
\Phi_0 \equiv p_z \,,\\
\Phi_{1r} \equiv P_r - p_r \,,\\
\Phi_{1\theta} \equiv P_\theta - p_\theta\,,\\
\Phi_{2r} \equiv \Pi_r \,,\\
\Phi_{2\theta} \equiv \Pi_\theta\,.
\end{cases}
\end{equation}
The canonical Hamiltonian is given by
\begin{equation}\label{HNPW}
H = \frac{p_\theta^2}{2mr^2}-z(r-a)
\end{equation}
and the demand of time conservation of the primary constraints (\ref{primaryconsts}) through the Dirac-Bergmann algorithm leads to a secondary one given by
\begin{equation}\label{r-a}
\Phi \equiv T(r,\theta) = r - a \,.
\end{equation}
Due to the discussion in the previous section we substitute the last constraint (\ref{r-a}) by the linear combination
\begin{equation}
\tilde{\Phi} = r - a - \Pi_r
\end{equation}
and the redefined constraints matrix is given by
\begin{equation}
C_{rs} = \left(
\begin{array}{cccccc}
0&0&0&0&0&0\\
0&0&0&-1&0&0\\
0&0&0&0&-1&0\\
0&1&0&0&0&0\\
0&0&1&0&0&0\\
0&0&0&0&0&0
\end{array}
\right)
\,.
\end{equation}
The two 
completely null rows above confirm the already stated fact that the model possesses two first class constraints, namely $\Phi_0$ and $\tilde{\Phi}$.
Therefore the singular Lagrangian (\ref{Linvrtheta}) constitutes a genuine first class system with gauge freedom equivalent to the original second class (\ref{Lrr}).

\subsection*{Elliptical Path}
As a second interesting example we consider a particle in a plane confined to move along the ellipse
\begin{equation}
(x/a)^2+(y/b)^2 = 1
\,.
\end{equation}
The two fixed real numbers $a$ and $b$ denote the major and minor axes.  In principle this constraint could be treated as a particular case of a specific conic of the form
\begin{equation}\label{oldT}
 T(x,y) = \frac{1}{2} Ax^2 + \frac{1}{2} B y^2 + Cxy +Dx +Ey +F = 0
\end{equation}
as described in the references \cite{Barbosa:2018bng, Barbosa:2018dmb}.  However,
in order to compare the results with the rigid rotor as discussed in the previous
example and with the gauge invariant system of Nemeschansky, Preitschopf and Weinstein \cite{Nemeschansky:1987xb} for the special case of a degenerated ellipse when $b=a$, we include a square root in the constraint equation and
consider instead the starting Lagrangian\footnote{This choice explicitly shows that the present generalization is not restricted to the form (\ref{oldT}).}
\begin{equation}\label{Lel}
L = \frac{m(\dot{x}^2+\dot{y}^2)}{2}
+z\sqrt{ab}\left[
\sqrt{(x/a)^2+(y/b)^2}-1
\right]
\,.
\end{equation}
Note that for $b=a$ this reproduces back precisely the Lagrangian (\ref{Lrr}) in polar coordinates.
Associated to (\ref{Lel}) we have the canonical Hamiltonian
\begin{equation}\label{Hel}
H = \frac{p_x^2+p_y^2}{2m} - z\left[ {\cal B} - \sqrt{ab} \right]
\end{equation}
and the four second class constraints
\begin{equation}\label{72}
\begin{cases}
\chi_1\equiv p_z\,,\\
\chi_2\equiv T(x,y) \equiv {\cal B}-\sqrt{ab}\,,\\
\chi_3\equiv \displaystyle\frac{xp_xb/a+yp_ya/b}{m{\cal B}}\,,\\
\chi_4\equiv \displaystyle\frac{(xp_y-yp_x)^2}{m^2{\cal B}^3}+\frac{z(x^2b^2/a^2+y^2a^2/b^2)}{m{\cal B}^2}\,,
\end{cases}
\end{equation}
where, for convenience, we have defined
\begin{equation}
{\cal B}\equiv\sqrt{x^2b/a+y^2a/b}
\,.
\end{equation}
The Hamiltonian (\ref{Hel}) and constraint set (\ref{72})
agree with equations (\ref{H}), (\ref{chi3}) and (\ref{chi4})
for $n=2$, $q_1=x$, $q_2=y$, $q_0=z$, $f_{ij}=\delta_{ij}$ and $Q_{ij}$ given by
\begin{equation}
Q_{ij}
=
\displaystyle
\frac{1}{{\cal B}^3}
\left[
\begin{array}{cc}
y^2 & -xy \\
-xy & x^2  
\end{array}
\right]
\,.
\end{equation} 
The non-null Dirac brackets can be read directly from equations (\ref{qipj}) to (\ref{q0pi}) as
\begin{equation}
[x,p_x] = \frac{y^2a^4}{x^2b^4+y^2a^4}
\,,~~~~~[y,p_y] = \frac{x^2b^4}{x^2b^4+y^2a^4}
\,,
\end{equation}
\begin{equation}
[x,p_y] = -\frac{xya^2b^2}{x^2b^4+y^2a^4} = [y,p_x]
\,,
\end{equation}
\begin{equation}
[p_x,p_y] = \frac{(yp_x-xp_y)a^2b^2}{x^2b^4+y^2a^4}
\,,
\end{equation}
\begin{equation}
[z,x]=\frac{2y(yp_x-xp_y)a^3b}{m(x^2b^4+y^2a^4){\cal B}}
\,,
~~~~~
[z,y]=\frac{2x(xp_y-yp_x)ab^3}{m(x^2b^4+y^2a^4){\cal B}}
\,,
\end{equation}
\begin{eqnarray}
[z,p_x]&=&\frac{2}{{\cal D}^2{\cal B}^3}
\left\lbrace
\frac{p_y^2}{m}
\left[ -2x^3y^2a/b - x^5b/a -xy^4a^3/b^3 \right]
\right.\nonumber\\&&\left.
~~~~+\frac{p_xp_y}{m}
\left[ 2x^2y^3 a/b + x^4y b/a +y^5a^3/b^3 \right] 
\right.
\nonumber\\&&\left.
\phantom{\frac{A}{B}}+z{\cal B}^2\left[ 
-x^3y^2b/a+xy^4a^3/b^3-xy^4a/b+x^3y^2a/b
\right]
\right\rbrace
\,,
\end{eqnarray}
and
\begin{eqnarray}
[z,p_y]&=&\frac{2}{{\cal D}^2{\cal B}^3}
\left\lbrace
\frac{p_x^2}{m}
\left[ -2x^2y^3b/a -x^4ya^3/b^3 -y^5a/b \right]
\right.\nonumber\\&&\left.
~~~~
+\frac{p_xp_y}{m}
\left[ 2x^3yb/a + xy^4 a/b + x^5b^3/a^3 \right] 
\right.
\nonumber\\&&\left.
\phantom{\frac{A}{B}}+z{\cal B}^2\left[ 
x^2y^3b/a+x^4yb^3/a^3-x^2y^3a/b-x^4yb/a
\right]
\right\rbrace
\end{eqnarray}
with
\begin{equation}
{\cal D}\equiv x^2b^2/a^2 + y^2a^2/b^2
\,.
\end{equation}

As we have seen,
it is possible to obtain a gauge invariant version associated to the second class system (\ref{Lel}).  In fact, in the present case, the potential (\ref{W}) reads
\begin{equation}
W(x,y,p_x,p_y)=\frac{(xp_yb/a-yp_xa/b)^2}{2m(x^2b^2/a^2+y^2a^2/b^2)}
\end{equation}
leading to the first-order Lagrangian
\begin{equation}
L_{fo} = p_x\dot{x}+p_y\dot{y}
-\frac{(xp_yb/a-yp_xa/b)^2}{2m(x^2b^2/a^2+y^2a^2/b^2)}
+z\sqrt{ab}\left[
\sqrt{(x/a)^2+(y/b)^2}-1
\right]
\end{equation}
whose corresponding action is gauge invariant under
\begin{equation}
\begin{cases}
z\rightarrow \dot{\Omega}\,, \\
p_x \rightarrow p_x + \displaystyle\frac{bx \Omega}{a{\cal B}}\,, \\
p_y\rightarrow p_y + \displaystyle\frac{ay \Omega}{b{\cal B}}\,,
\end{cases}
\end{equation}
for an arbitrary time-dependent function $\Omega$.
Note that for the particular case $b=a$, corresponding to a degenerate ellipse, we recover the previous rigid rotor example in Cartesian coordinates.

\section{Conclusion and Final Remarks}
The study of simple crafty mechanical systems can shed light into more involved field theory models with similar inner structure regarding, for instance, the constraint algebra, gauge and BRST symmetries, Grassmann variables and quantization aspects.  The quantum rigid rotor can be described either as a second class system or as a particular gauge invariant model where its BRST rich structure has been explored in
\cite{Nemeschansky:1987xb, Gupta:2009dy, Shukla:2014hea, Shukla:2014spa}.  After the partial generalization
of the main ingredients present in that model enabling gauge symmetry to conic constraints in references \cite{Barbosa:2018bng, Barbosa:2018dmb}, we have seen in the present paper that it is possible to push
it considerably further and shown that the quantum rigid rotor constitutes a very special case of a broader class of constrained systems of the form (\ref{L}), all of which can be made gauge invariant as described
by Lagrangian (\ref{Lfo}).
The trivial circular symmetry of a 2D rigid rotor is not a {\it sine qua non} condition for the gauge invariance of that model, but an accidental particular case of the deeper framework which we have discussed.
A careful study of these previous works on the rigid rotor gauge symmetry have then provided a motivation to obtain a general method for converting second class systems of this form to first class ones.  By means of writing a first-order Lagrangian and using the fine tuned potential (\ref{W}) it was possible to produce the gauge symmetry (\ref{gs}) without changing the physical content of the theory.  
After discussing the general framework and establishing our main results, 
we have explicitly shown the details of the method producing the corresponding gauge invariant action for the examples of the quantum rigid rotor and a particle constrained to move on an elliptical path.  Therefore we have reproduced the well known gauge invariant Lagrangian of the rigid rotor from scratch in a clear and systematic way.  In particular, for the elliptical path case, we have used a square root constraint type,
stressing the strength and generality of the method. In that case it is clear that it is the geometrical constraint surface which dictates the physics, rather than its algebraic description.
Further generalization of the method described in this letter as well as its application to quantum field models is currently under investigation.

\end{document}